\documentclass{article}

\usepackage{arxiv}

\usepackage[utf8]{inputenc} % allow utf-8 input
\usepackage[T1]{fontenc}    % use 8-bit T1 fonts
\usepackage{hyperref}       % hyperlinks
\usepackage{url}            % simple URL typesetting
\usepackage{booktabs}       % professional-quality tables
\usepackage{amsfonts}       % blackboard math symbols
\usepackage{nicefrac}       % compact symbols for 1/2, etc.
\usepackage{microtype}      % microtypography
\usepackage{graphicx}
\usepackage{natbib}
\usepackage{doi}
\usepackage{threeparttable}
\usepackage{rotating}
\usepackage{caption}
\usepackage{amsmath}
\usepackage{multirow}
\usepackage{bm}
\usepackage{array}

\def\bX{{\bf X}}

\def\b0{{\bf 0}}

\def\balpha{\mbox{\boldmath $\alpha$}}

\def\bbeta{\mbox{\boldmath $\beta$}}

\title{A Phylogeny-based Test of Mediation Effect in Microbiome}

\author{ {Qilin Hong}\\
	Department of Biostatistics and Medical Informatics\\
	University of Wisconsin-Madison\\
	Madison, WI 53715, U.S.A. \\
	\And
	{Guanhua Chen} \\
    Department of Biostatistics and Medical Informatics\\
	University of Wisconsin-Madison\\
	Madison, WI 53715, U.S.A. \\
	\texttt{gchen25@wisc.edu} \\
	\And
	{Zheng-Zheng Tang} \\
    $^{1}$ Department of Biostatistics and Medical Informatics\\
	University of Wisconsin-Madison\\
	Madison, WI 53715, U.S.A. \\
	$^{2}$ Wisconsin Institute for Discovery\\
	Madison, WI 53715, U.S.A. \\
	\texttt{tang@biostat.wisc.edu} \\
}

\hypersetup{
pdftitle={A Phylogeny-based Test of Mediation Effect in Microbiome},
pdfsubject={Mathematics},
pdfauthor={Qilin Hong, Guanhua Chen, Zheng-Zheng Tang},
pdfkeywords={Composite null hypothesis, Joint significance test, Mediation analysis, Microbiome, Phylogenetic tree},
}

\begin{document}
\maketitle

\begin{abstract}
	Recent studies suggest that the microbiome can be an important mediator in the effect of a treatment on an outcome. Microbiome data generated from sequencing experiments contain the relative abundance of a large number of microbial taxa with their evolutionary relationships represented by a phylogenetic tree. The compositional and high-dimensional nature of the microbiome mediator invalidates standard mediation analyses. We propose a phylogeny-based mediation analysis method (PhyloMed) for the microbiome mediator.  PhyloMed models the microbiome mediation effect through a cascade of independent local mediation models on the internal nodes of the phylogenetic tree. Each local model captures the mediation effect of a subcomposition at a given taxonomic resolution. The method improves the power of the mediation test by enriching weak and sparse signals across mediating taxa that tend to cluster on the tree. In each local model, we further empower PhyloMed by using a mixture distribution to obtain the subcomposition mediation test $p$-value, which takes into account the composite nature of the null hypothesis. PhyloMed enables us to test the overall mediation effect of the entire microbial community and pinpoint internal nodes with significant subcomposition mediation effects.  Our extensive simulations demonstrate the validity of PhyloMed and its substantial power gain over existing methods. An application to a real study further showcases the advantages of our method.
\end{abstract}

% keywords can be removed

\section{Introduction}
\label{s:intro}
Recent microbiome studies have revealed important associations between microbiome and treatment responses \citep{koeth2013intestinal,zeevi2015personalized, taur2016microbiome, kuntz2017introducing}. For example, gut microbes are found to greatly influence the potency of immunotherapy and some chemotherapies with immunostimulatory functions in the treatments for cancer \citep{fessler2019exploring}. 
These studies suggest that the human microbiome is emerging as a crucial mediator between the treatment and its outcomes. Finding the mediating microbial taxa and elucidating the mechanism behind the mediation can greatly facilitate the development of strategies to manipulate the microbiome to augment therapeutic responses.

Mediation analysis provides the statistical framework to investigate if a treatment or exposure affects an outcome through a mediator. 
%The potential outcome framework to establish of the causal effect interpretation. 
%Donald Rubin and Paul Holland developed the potential outcome
%idea (Neyman, 1923) and established a formal mathematical
%causal framework for both observational and randomized experimental
%studies (Holland, 1986; Rubin, 1974, 2005).
The traditional meditation analysis studies the mediation effect of a single mediator \citep{baron1986moderator, mackinnon2012introduction}. More recent developments have extended that to multiple and even high-dimensional mediators \citep{vanderweele2014mediation, daniel2015causal, zhang2016estimating, huang2016hypothesis, huang2019variance}.
%\citep{vanderweele2015explanation}

Microbiome data from sequencing experiments contain the relative abundance of a large number of different microbes. The compositional and high-dimensional nature of the microbiome mediators poses a great challenge to standard mediation analyses.  
Distance-based mediation tests MedTest \citep{zhang2018distance} and MODIMA  \citep{hamidi2019modima}  have been developed for testing the mediation effect of the entire microbial community. These global tests can achieve good power when many microbial taxa in the community mediate the treatment effect on the outcome. However, their power is limited when the number of mediating taxa is small (i.e. sparse mediation signal). Moreover, this type of method is not designed to identify mediating taxa. 

Another mediation analysis methods for microbiome data assume sparse mediation effects and estimate the effects via regularization
%, such as CMM (compositional causal
%mediation model) \citep{sohn2019compositional} and SparseMCMM (sparse microbial causal mediation model) \citep{wang2020estimating},
%are based on the regularized mediation effect estimation, with the assumption of sparse mediation effects 
\citep{sohn2019compositional, wang2020estimating, zhang2021mediation}. These methods aim to select mediating taxa and some provide global tests of the overall mediation effect at the community level. These global tests are potentially more powerful than the distance-based tests in the scenario of sparse mediation. However, these methods need to model the high-dimensional microbial composition, which is a daunting task due to the complex structure of microbiome data such as compositionality, over-dispersion, and high zero-inflation. Misspecification of the model can lead to biased estimates and tests of mediation effects. 

In this paper, we develop a new phylogeny-based mediation analysis method (PhyloMed) for high-dimensional microbial composition mediators.  
Microbial taxa are evolutionarily related and their relationship is represented by a phylogenetic tree.
Incorporating phylogenetic information into the analysis of microbiome data has been shown to greatly improve the performance of a variety of statistical analyses \citep{ washburne2018methods}. 
Phylogenetic distances such as UniFrac \citep{lozupone2005unifrac} are useful in improving the power of association testing and the accuracy of sample classification.
Another common way of utilizing the tree information is to break down full microbial composition according to the tree structure, which allows more flexible modeling of microbiome data and locating the signals to a certain taxonomic
level \citep{tang2017general, wang2017dirichlet}. This idea motivates the development of PhyloMed.
%While there has been a recent expansion of phylogenetically informed analytical tools, 
%To the best of our knowledge, 
%However, besides the aforementioned distance-based methods, no phylogeny-aware mediation analysis methods for microbiome are currently available.
%In this paper, we introduce a phylogeny-based mediation test (PhyloMed) for high-dimensional microbial composition mediators. 
PhyloMed models the microbiome mediation effect through a cascade of independent local mediation models of subcompositions on the internal nodes of the phylogenetic tree. 
Hence, PhyloMed avoids the difficulty of directly modeling the high-dimensional full composition. 
Depending on the depth of internal nodes on the tree, these subcompositions represent the relative abundance of groups of taxa at various taxonomic resolutions. Descendants of each internal node share a certain degree of evolutionary affinity and tend to have similar biological functions. Therefore, mediating taxa are likely to cluster on the tree and PhyloMed can potentially enrich mediation signal and improve the power of mediation analysis.

In each local model of PhyloMed, we test the mediation effect of a subcomposition in the treatment $\rightarrow$ outcome pathway.
The mediation effect is commonly expressed as a product of two parameters, the treatment-mediator association and the mediator-outcome association conditional on the treatment.
The null hypothesis of no mediation effect is composite: either one of those associations is zero or both are zeros.
Traditional tests such as Sobel's test \citep{sobel1982asymptotic} and joint significance test are overly conservative and yield low power because they ignore the composite nature of the null hypothesis \citep{mackinnon2002comparison, barfield2017testing}. 
To address this problem, we estimate the proportions of component null hypotheses among all the local mediation models and obtain the subcomposition mediation test $p$-value using a mixture distribution with three components, each of which corresponds to one type of null hypothesis. 
This indicates another advantage of decomposing the full composition mediation model into multiple subcomposition local models: a large number of local tests provides the opportunity to accurately estimate the proportions of different nulls so that the mediation tests based on the mixture distribution have the proper size. 
%testing subcomposition mediation effect on the tree -- the multiple numbers of testing enable us to accurately estimate the proportions of different nulls. 
%The resulting test has better type I error control and substantial increased power than the traditional tests.

By combining all the subcomposition mediation test $p$-values across nodes of the phylogenetic tree, PhyloMed produces a global mediation test to evaluate the mediation signal at the microbial community level. 
Simulation studies show that the PhyloMed global test 
better controls the type I error and
is substantially more powerful than existing global mediation tests in the presence of the sparse signal.
In addition, by detecting significant subcomposition mediation test $p$-values, PhyloMed can pinpoint the nodes of the tree with groups of descendant taxa that are potential mediators. 
Simulation studies show that PhyloMed properly controls false discovery rate (FDR) in identifying mediating nodes and yields higher discovery power than the procedures that use Sobel's and joint significance tests to obtain subcomposition mediation $p$-values. 
%apply the method to a real microbiome dataset to investigate
%an effect of fat intake on body mass index mediated through the gut
%microbiome.
In an application to a real microbiome  study, PhyloMed reveals that the effect of antibiotics treatment on weight gain
is transmitted through perturbing the gut microbial community and highlights mediating subcompositions on the phylogenetic tree.

\section{Methods}
\label{s:methods}

\subsection{Phylogeny-based mediation model}

\begin{figure}[htp]
	\centering
	\includegraphics[scale=0.7]{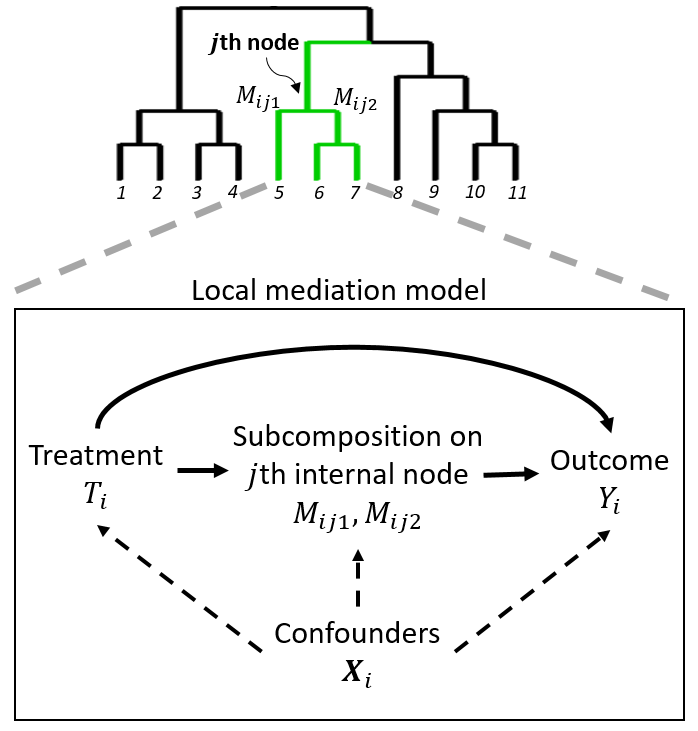}
	\caption{Example phylogenetic tree and causal path diagram of a local mediation model. PhyloMed divides the high-dimensional microbial composition into multiple subcompositions at internal nodes of the phylogenetic tree and tests the mediation effect of each subcomposition. 
		%We aim to investigate the effect of treatment directly on the outcome (i.e.  direct effect) and the effect of treatment through microbiome composition as mediators (i.e. mediation effect), while adjusting for a set of pre-treatment confounders
	}
	\label{fig:mediation}
\end{figure}

The phylogeny is an estimation of the microorganisms' evolutionary history.
%and classifies every organism by a series of splits corresponding to estimated events in which a most recent common ancestor speciated to form two daughter species.
%For ease of notation, we assume there is one covariate. 
Figure \ref{fig:mediation} shows an example of a simple rooted binary phylogenetic tree with 11 microbial taxa at terminal nodes and 10 internal nodes, representing the common ancestors of those taxa. The subcomposition on a given internal node consists of the relative abundance aggregated at its two child nodes. For example, the highlighted internal node in Figure \ref{fig:mediation} has a terminal node for taxon 5 as its left child node and the most recent common ancestor of taxa 6 and 7 as its right child node. Therefore, the subcomposition defined on that internal node has two components: one is the abundance of taxon 5 and the other is the aggregated abundance of taxa 6 and 7. 

As depicted in Figure \ref{fig:mediation}, we propose to construct a local mediation model for the subcomposition on each internal node of the phylogenetic tree. 
Suppose there are $J$ internal nodes on the tree
and data contain a random sample of $n$ subjects from a population. For internal node $j=1, \ldots, J$ and subject $i=1, \ldots, n$, we let $(M_{ij1}, M_{ij2})$ be the subcomposition mediator, $\mathbb{N}_{ij}=M_{ij1}+M_{ij2}$ be the counts assigned to the $j$th internal node of the tree, $T_i$ be the treatment variable, $Y_i$ be the outcome variable, and $\bX_i$ be a set of confounders that may affect the treatment, mediator, and outcome. 

	%We assume that the microbiome data are generated hierarchically, conditioning on the total read count of each internal node. 
Our method separately models subcomposition  $(M_{ij1}, M_{ij2})$ conditional on $\mathbb{N}_{ij}$ for each internal node $j$ and assumes $\log(M_{ij1}/M_{ij2})$'s over internal nodes are mutually independent normal random variables.
%Specifically, we assumed $\log(M_{ij1}/M_{ij2})$'s over internal nodes are mutually independent normal random variables.}
%We need to convert the two-part composition $(M_{ij1}, M_{ij2})$ to a variable in the Euclidean space
To represent the causal path diagram of the local mediation model at the $j$th internal node, we apply the following linear and generalized linear regression models
\begin{equation}\label{Mmodel}
E\left\{ \log \left( \frac{M_{ij1}}{M_{ij2}}\right) \right\} = \balpha^{\rm T}_{jX} \bX_i + \alpha_j T_i,
\end{equation}
\begin{equation}\label{Ymodel}
g\{E(Y_i)\} = \bbeta^{\rm T}_{jX} \bX_i + \beta_{jT} T_i  + \beta_{j} \log\left(\frac{M_{ij1}}{M_{ij2}}\right),
\end{equation}
where $g\{\cdot\}$ is the link function depending on the type of the outcome and we omit the intercept term in both models as it can be absorbed into $\bX_i$. 
%The (\ref{Mmodel}) is a linear model and (\ref{Ymodel}) is a generalized linear model.
In both models, we use the log ratio of $M_{ij1}$ to $M_{ij2}$ as the mediator variable, which is a common scale-invariant transformation for compositions. %\red{and independence of $\mathbb{N}_{ij}$}.   
Because $\log(M_{ij1}/M_{ij2})$'s over internal nodes are independent, rather than fitting the large joint model $[Y_i \mid \bX_i, T_i, \log(M_{i11}/M_{i12}), \ldots, \log(M_{iJ1}/M_{iJ2})]$, we can fit $J$ low-dimensional models $[Y_i \mid \bX_i, T_i, \log(M_{ij1}/M_{ij2})]$ as equation  (\ref{Ymodel}) for the purpose of hypothesis testing.

The potential outcome framework \citep{robins1992identifiability, vanderweele2009conceptual, imai2010identification} has established a series of identifiability assumptions such that models (\ref{Mmodel}) and (\ref{Ymodel}) lead to quantification of the causal mediation effect. A rigorous definition of causal mediation using potential outcomes framework is provided in Web Appendix A.1. 
An extension of the model to allow for the treatment-mediator interaction is described in the Web Appendix A.2. 

Our method concerns the mediation effects at internal nodes, rather than leaf nodes of a tree. Note that the meaning of ``mediation effect'' at internal nodes and leaf nodes are not exactly equivalent. Web Figure S1 uses a simple example to show different scenarios of mediation null and alternative hypotheses at the leaves and at the upper level after aggregating leaf-level taxa. The presence/absence of the mediation effect at the upper-level nodes corresponds to that at the leaf nodes, except for the situation shown in the last column of the figure. In that scenario, the set of treatment-associated taxa and the set of outcome-associated taxa are completely separate. After aggregating these taxa at their common ancestors, the upper-level nodes will have the mediation effect, though the treatment and outcome associations are contributed by different descendants. Therefore, if we want to interpret the significant test results on the internal node down to the leaf level, we need to assume that at least one leaf-level descendant is associated with both the treatment and the outcome. This assumption is mild and unavoidable if one wants to perform analysis at any aggregated taxa.

\subsection{Composite null hypothesis tests in local mediation models}
In each local mediation model, we are interested in testing whether the subcomposition at the $j$th internal node lies in the causal pathway from the treatment to the outcome.  
The null and alternative hypotheses for this testing problem can be formulated as 
\begin{equation}
H^{j}_{0}: \alpha_j \beta_{j}=0 \quad {\rm vs} \quad H^{j}_{a}: \alpha_j \beta_{j}\neq 0.
\end{equation}
The null hypothesis can be equivalently expressed as the union of three disjoint component null hypotheses
\begin{align*} 
H^{j}_{00}&: \alpha_j = \beta_{j} = 0,\\
H^{j}_{10}&: \alpha_j \neq 0, \beta_{j} = 0,\\
H^{j}_{01}&: \alpha_j = 0, \beta_{j} \neq 0.
\end{align*} 

Sobel's test \citep{sobel1982asymptotic} and joint significance test have been widely applied to test the mediation null hypothesis. Unfortunately, these tests have severely deflated type I error and lack power because they fail to take into account the composite nature of the null hypothesis \citep{mackinnon2002comparison, barfield2017testing}. 
 
To address this issue, several new mediation tests  were recently developed \citep{huang2019genome, dai2020multiple, liu2021large}. 
In light of these methods, we propose a new testing procedure that handles the composite null hypothesis and produces well-controlled type I error for multiple local mediation tests.
%and is suitable for general use in microbiome mediation studies.
Let $P_{\alpha_j}$ and $P_{\beta_j}$ denote the $p$-values for testing $\alpha_j=0$ and $\beta_{j}=0$, respectively. 
The $P_{\alpha_j}$ and $P_{\beta_j}$ are asymptotically uniformly distributed under their respective null hypotheses and are independent under the no unmeasured confounding assumptions.
%many microbiome studies have very small sample size, which. the asymptotic test may not be accurate when sample size is small
In PhyloMed,  
we adopt score statistics and obtain the observed $p$-values $p_{\alpha_j}$ and $p_{\beta_j}$ based on the reference asymptotic distribution or permutation (details in Web Appendix B). 
The permutation $p$-value is more accurate than its asymptotic counterpart when the study sample size is small, which is usually the case for microbiome studies of treatment response.   
We implement an adaptive procedure to efficiently and accurately obtain the permutation $p$-values (Web Appendix B.3).

We then define the mediation test statistic for $H^j_0$ as
\begin{equation}
P_{\max_j} = \max(P_{\alpha_j}, P_{\beta_j}).
\end{equation}
The classical joint significant test also takes $P_{\max_j}$ as the mediation test statistic and determines statistical significance using the uniform distribution. 
In fact, $P_{\max_j}$ follows a mixture distribution with three components, each of which corresponds to one type of null hypothesis $H^j_{00}$, $H^j_{10}$, or $H^j_{01}$. 
Let $p_{\max_j}$ denote $\max(p_{\alpha_j}, p_{\beta_j})$, and $\pi_{00}$, $\pi_{10}$, and $\pi_{01}$ be the probabilities of the three null hypotheses among the $J$ local mediation models defined on the tree.  The $p$-value of mediation test in the $j$th local model is given by
%Under the null hypothesis of no mediation effect in the $j$th local model, we have
\begin{align}\label{medpval}
& Pr(P_{\max_j} \leq p_{\max_j})  \nonumber\\
%= &Pr(p_{\max_j} \leq t \mid H^j_{00}) Pr(H^j_{00}) + Pr(p_{\max_j} \leq t \mid H^j_{10}) Pr(H^j_{10}) + Pr(p_{\max_j} \leq t \mid H^j_{01}) Pr(H^j_{01}) \nonumber\\
= &\pi_{00} Pr(P_{\alpha_j} \leq p_{\max_j}, P_{\beta_j} \leq p_{\max_j} \mid H^j_{00})  \nonumber\\
 &+ \pi_{10} Pr(P_{\alpha_j} \leq p_{\max_j}, P_{\beta_j} \leq p_{\max_j} \mid H^j_{10}) + \pi_{01} Pr(P_{\alpha_j} \leq p_{\max_j}, P_{\beta_j} \leq p_{\max_j} \mid H^j_{01})  \nonumber\\
= &\pi_{00} Pr(P_{\alpha_j} \leq p_{\max_j} \mid \alpha_j = 0) Pr( P_{\beta_j} \leq p_{\max_j} \mid \beta_j=0)  \nonumber\\
&+ \pi_{10} Pr( P_{\beta_j} \leq p_{\max_j} \mid \beta_j=0) Pr(P_{\alpha_j} \leq p_{\max_j} \mid \alpha_j \neq 0)  \nonumber\\
&+ \pi_{01} Pr(P_{\alpha_j} \leq p_{\max_j} \mid \alpha_j = 0) Pr( P_{\beta_j} \leq p_{\max_j} \mid \beta_j \neq 0) \nonumber\\
= &\pi_{00} p^2_{\max_j} + \pi_{10} p_{\max_j} Pr(P_{\alpha_j} \leq  p_{\max_j} \mid \alpha_j \neq 0) + \pi_{01} p_{\max_j} Pr(P_{\beta_j} \leq p_{\max_j} \mid \beta_j \neq 0).
\end{align}
In this formula, we need to estimate three probabilities: $\pi_{00}$, $\pi_{10}$, and $\pi_{01}$, and two power functions evaluated at $p_{\max_j}$: $Pr(P_{\alpha_j} \leq p_{\max_j} \mid \alpha_j \neq 0)$ and $Pr(P_{\beta_j} \leq p_{\max_j} \mid \beta_j \neq 0)$.

We first employ the method proposed by \citet{jin2007estimating}
%Storey's method \citep{storey2002direct, storey2004strong} 
to estimate ${\pi}_{0\bullet}$, the proportion of null $\alpha_j=0$, using $p_{\alpha_j}$'s in all $J$ local mediation models. Specifically, we convert the $p_{\alpha_j}$ to $Z$-score 
$
Z_{\alpha_j} = sign(\widehat\alpha_j) \times \Phi^{-1}(1-p_{\alpha_j}/2),
$
where $sign(\widehat\alpha_j)$ is the sign of the $\alpha_j$ estimate and $\Phi^{-1}(x)$ is the quantile function of the standard normal. 
\citet{jin2007estimating} use the empirical characteristic function and Fourier analysis to estimate the proportion of nulls. The empirical characteristic function is defined as
\begin{equation*}
\varphi_J(t; Z_{\alpha_1}, \ldots, Z_{\alpha_J}) = \frac{1}{J}\sum_{j=1}^J e^{itZ_{\alpha_j}},
\end{equation*}
where $i=\sqrt{-1}$. The proportion of nulls can be consistently estimated as
\begin{equation*}
\widehat{\pi}_{0\bullet} = \inf_{\left\{0 \leq t \leq \sqrt{\log(J)}\right\}} \left[ \int_{-1}^{1} \left(1-|\xi|\right)\left\{{\rm Re}(\varphi_J(t\xi; Z_{\alpha_1}, \ldots, Z_{\alpha_J})) e^{t^2 \xi^2/2}\right\} \, d\xi \right],
\end{equation*}
where ${\rm Re}(x)$ denotes the real part of a complex number $x$.
Similarly, we can obtain $\widehat{\pi}_{\bullet0}$, the estimated proportion of null $\beta_j=0$ using $p_{\beta_j}$'s across all local models.
%Let $\widehat{\pi}_{0\bullet}$ and $\widehat{\pi}_{\bullet0}$ denote these two proportion estimates, 
Then, under $H_0^j$, the estimates of $\pi_{00}$, $\pi_{10}$, $\pi_{01}$ are  $\widehat{\pi}_{00}=\widehat{\pi}_{0\bullet}\widehat{\pi}_{\bullet0}/\widehat{\pi}_0$, $\widehat{\pi}_{10}=(1-\widehat{\pi}_{0\bullet})\widehat{\pi}_{\bullet0}/\widehat{\pi}_0$, and
$\widehat{\pi}_{01}=\widehat{\pi}_{0\bullet}(1-\widehat{\pi}_{\bullet0})/\widehat{\pi}_0$, where $\widehat{\pi}_0=\widehat{\pi}_{0\bullet} + \widehat{\pi}_{\bullet0}-\widehat{\pi}_{0\bullet}\widehat{\pi}_{\bullet0}$.

For the two power functions, we employ the nonparametric estimates based on the Grenander estimator of the $p$-value density \citep{langaas2005estimating}. We show below the procedure of obtaining $\widehat{Pr}(P_{\alpha_j} \leq p_{\max_j} \mid \alpha_j \neq 0)$.
Specifically, we use formula (9) in \cite{langaas2005estimating} to compute the Grenander estimator of the decreasing density at every $p_{\alpha_j}$, denoted by $\widehat{f}(p_{\alpha_j})$. 
Under $H_0^j$, $P_{\alpha_j}$ follows a mixture distribution composed of the uniform distribution (under $H^j_{00}$ and $H^j_{01}$) and the power function (under $H^j_{10}$). Therefore, we can obtain the conditional density estimate $\widehat{f}(p_{\alpha_j} | \alpha_j \neq 0)$ by solving the equation 
\begin{equation*}
\widehat{f}(p_{\alpha_j}) = \widehat{\pi}_{10} \widehat{f}(p_{\alpha_j} \mid \alpha_j \neq 0) + \widehat{\pi}_{00} + \widehat{\pi}_{01}.
\end{equation*}
We can then estimate $\widehat{Pr}(P_{\alpha_j} \leq p_{\max_j} \mid \alpha_j \neq 0)$ based on $\widehat{f}(p_{\alpha_j} \mid \alpha_j \neq 0)$ at all observed $p_{\alpha_j}$'s.
Using the similar estimation procedure, we can obtain $\widehat{Pr}(P_{\beta_j} \leq p_{\max_j} \mid \beta_j \neq 0)$. 
Finally, the mediation test $p$-value at the $j$th node can be estimated as $p_j = \widehat{\pi}_{00} p^2_{\max_j} + \widehat{\pi}_{10} p_{\max_j} \widehat{Pr}(P_{\alpha_j} \leq  p_{\max_j} \mid \alpha_j \neq 0) + \widehat{\pi}_{01} p_{\max_j} \widehat{Pr}(P_{\beta_j} \leq p_{\max_j} \mid \beta_j \neq 0)$.

\subsection{Global mediation test and mediating node detection}
Under the global null mediation hypothesis that there is no mediation effect in any of the internal nodes (i.e. $H_0: \cap_{j=1}^J H^j_0$ ), we show below that the mediation test statistics $P_{\max_j}$'s across all internal nodes are asymptotically mutually independent as the sample size goes to infinity.

First of all, under $H^j_0$, the two power function probabilities $Pr(P_{\alpha_j} \leq t_j \mid \alpha_j \neq 0)$ and $Pr(P_{\beta_j} \leq t_j \mid \beta_j \neq 0)$ converge to 1 for a constant $t_j$ when sample size goes to infinity. Therefore, based on formula (\ref{medpval}), $Pr(P_{\max_j} \leq t_j)$ can be approximated by
%Based on formula (\ref{medpval}), under $H^j_0$, 
$\pi_{00} Pr(P_{\alpha_j} \leq t_j \mid \alpha_j = 0) Pr( P_{\beta_j} \leq t_j \mid \beta_j=0)  + \pi_{10} Pr( P_{\beta_j} \leq t_j \mid \beta_j=0) + \pi_{01} Pr(P_{\alpha_j} \leq t_j \mid \alpha_j = 0) =
\pi_{00} t^2_{j} + \pi_{10} t_{j} + \pi_{01} t_{j}$.
Clearly, the formula only involves probabilities of $P_{\alpha_j}$ and $P_{\beta_j}$ under their respective nulls $\alpha_j=0$ and $\beta_j = 0$.
%Due to the independence of $\log(M_{ij1}/M_{ij2})$'s over internal nodes, all p-values of $P_{\beta_j}$'s under $\beta_j = 0$ are independent and all p-values of $P_{\alpha_j}$'s under $\alpha_j = 0$ are independent.}

Without loss of generality, we order the internal nodes $1, \ldots, J$ such that each parent node always appears in front of its children. 
Let $\mathbb{N}_J = \left\{\mathbb{N}_{iJ} \right\}^n_{i=1}$ be the vector of aggregated relative abundance at the $J$th node for all subjects. 
%As shown in the previous work on modeling subcompositions on a tree \citep{tang2017general, wang2017dirichlet, tang2018phylogenetic}, for any $j=1,\ldots,J$, conditional on $\mathbb{N}_j$, the subcomposition at the $j$th node is independent of the subcompositions at the nodes $1, \ldots, j-1$. Therefore, $P_{\max_J}$ is independent of all other $P_{\max_j}$'s conditional on $\mathbb{N}_J$ and 
Under the global null hypothesis $H_0$, we have 
\begin{align}
Pr\left\{\cup_{j=1}^J\left(P_{\max_j}\leq t_j\right)\right\} &= E\left[Pr\left\{\cup_{j=1}^J\left(P_{\max_j}\leq t_j\right) \mid \mathbb{N}_J\right\}\right] \nonumber \\
&\approx E\left[Pr\left(P_{\max_J}\leq t_J \mid \mathbb{N}_J\right)
Pr\left\{\cup_{j=1}^{J-1}\left(P_{\max_j}\leq t_j\right) \mid \mathbb{N}_J\right\}  \right]  \nonumber \\
%&\approx F(t_J) E\left(Pr(\cup_{j=1}^{J-1}\{P_{\max_j}\leq t_j\} \mid \mathbb{N}_J)\right)  \nonumber\\
&= Pr(P_{\max_J}\leq t_J) E\left[Pr\left\{\cup_{j=1}^{J-1}\left(P_{\max_j}\leq t_j\right) \mid \mathbb{N}_J\right\}\right] \nonumber\\
%&= F(t_J) Pr(\cup_{j=1}^{J-1}\{P_{\max_j}\leq t_j\}).
&= Pr(P_{\max_J}\leq t_J) Pr\left\{\cup_{j=1}^{J-1}\left(P_{\max_j}\leq t_j\right)\right\}, \nonumber
\end{align}
where $\approx$ represents that $P_{\max_J}$ is asymptotically independent of all other $P_{max_j}$'s conditional on $\mathbb{N}_J$. This is because, conditional on $\mathbb{N}_J$, $P_{\alpha_J}$ is asymptotically independent of all other $P_{\alpha_j}$'s and $P_{\beta_J}$ is asymptotically independent of all other $P_{\beta_j}$'s under the respective nulls $\alpha_J=0$ and $\beta_J = 0$.
The second last equation holds because the two regressions in the local mediation models are operated via the scale-invariant log-ratio of the subcomposition, hence,  $Pr(P_{\max_J}\leq t_J \mid \mathbb{N}_J) = Pr(P_{\max_J}\leq t_J)$.

Repeating the above procedure iteratively for $J-1$, $J-2, \ldots, 1$ gives
\begin{equation}
Pr\left\{\cup_{j=1}^J\left(P_{\max_j}\leq t_j\right)\right\} \approx \prod_{j=1}^J Pr(P_{\max_j}\leq t_j).
\end{equation}

%because $Pr(P_{\alpha_j} \leq  t_j \mid \alpha_j \neq 0) \rightarrow 1$ and $Pr(P_{\beta_j} \leq t_j \mid \beta_j \neq 0) \rightarrow 1$ when the sample size is large.
%As two regressions in the local mediation model are operated via the ratio of $M_{ij1}$ and $M_{ij2}$,  $P_{\max_j}$ is independent of $\mathbb{N}_j$, therefore, $Pr(P_{\max_j} \leq t_j \mid \mathbb{N}_j)=Pr(P_{\max_j} \leq t_j) \approx  F(t_j) $. 

As the tests on internal nodes are asymptotically independent, to control FDR in multiple testing, we can apply Benjamini-Hochberg (BH) procedure \citep{benjamini1995controlling} to identify a collection of nodes with significant mediation effects on the phylogenetic tree. 
To test the global mediation null hypothesis $H_0$, we apply Simes's method \citep{simes1986improved} to combine all the local mediation test $p$-values. 
%Under the null, the pcombined is asymptotically uniformly distributed. 
This global test focuses on a few smallest $p$-values and is powerful to detect the sparse mediation signal among subcompositions on the tree. 
%\red{We emphasize that the goal of the paper is for introducing a phylogeny-based testing procedure that generates uniformly distributed p-values under the mediation null hypothesis for subsequent multiplicity adjustment
%using existing methods, rather than developing a multiplicity adjustment method.}

\section{Simulation Studies}
\label{s:sim}

\subsection{Simulation strategy}
We performed extensive simulation studies to evaluate the type I error and power of the PhyloMed global mediation test. We considered using the asymptotic and permutation $p$-values of $p_{\alpha_j}$ and $p_{\beta_j}$ in PhyloMed and referred to these two versions as PhyloMed.A and PhyloMed.P. 
We compared their performance
with two distance-based tests MedTest \citep{zhang2018distance} and MODIMA \citep{hamidi2019modima}. 
In these distance-based methods, we considered Bray-Curtis, Jaccard, weighted and unweighted UniFrac distances. 
In addition, we considered a test based on the regularized mediation model CMM \citep{sohn2019compositional}.
We chose CMM as the representative of regularized methods because it provides a global test and has a relatively flexible mediator model for microbial full compositions. 
We used the normality-based and bootstrap methods for the CMM global test as described in their paper \citep{sohn2019compositional}. 

To resemble reality, we used a real microbiome dataset from \cite{zeevi2015personalized} as a basis for the simulation. 
The data contained microbiome samples from $900$ healthy subjects. In each round of the simulation, we randomly sampled $n=50$ or $200$ subjects out of the $900$ and divided them into two equal-sized treatment groups ($T_i = 0 {~~\rm or~~} 1$).
We used the top $100$ most abundant bacterial taxa and the associated phylogenetic tree with $99$ internal nodes. Let $\mathcal{S}_\alpha$ and $\mathcal{S}_\beta$ denote the set of treatment-associated and outcome-associated taxa, respectively, $\mathcal{S}$ be the union of the two sets, and $|\mathcal{S}|$ be the number of elements in $\mathcal{S}$.  Under the null of no mediation effect, these two sets of taxa do not overlap and we consider five combinations of $(|\mathcal{S}_\alpha|, |\mathcal{S}_\beta|)$ values: $(0, 0)$, $(3, 0)$, $(6, 0)$, $(0, 3)$, $(0, 6)$. 
Different values represent different mixtures of nulls $H_{00}$, $H_{10}$, and $H_{01}$:  $(0, 0)$ pertains to the setting that all local models are under $H_{00}$; $(3 {~\rm or~} 6, 0)$
pertains to the setting that some local models are under $H_{00}$ and others are under $H_{10}$; $(0, 3 {~\rm or~} 6)$ pertains to the setting that some local models are under $H_{00}$ and others are under $H_{01}$.
Under the alternative, we let the two sets of taxa completely or partially overlap. 
%and considered $|\mathcal{S}_\alpha|=|\mathcal{S}_\beta|=3$ or $|\mathcal{S}_\alpha|=|\mathcal{S}_\beta|=6$. 
%In all simulation settings, 
When $|\mathcal{S}|=3$, we randomly selected a clade with three descendant taxa and worked with the three taxa.
%the three taxa are all descendants of a randomly selected clade
when $|\mathcal{S}|=6$, we randomly selected two clades with three descendant taxa in each and worked with the six taxa. 
%the six taxa are the only descendants of two randomly selected clades, each of which has three descendants. 
Therefore, in the setting where taxa in $\mathcal{S}_\alpha$ and $\mathcal{S}_\beta$ are completely overlapped, we have $3 \text{ or } 6$ mediating taxa.
In the setting where taxa in $\mathcal{S}_\alpha$ and $\mathcal{S}_\beta$ are partially overlapped, we randomly selected one descendant under each clade as a mediating taxon such that we have 1 or 2 mediating taxa.

To perturb the abundance of each treatment-associated taxon $k\in \mathcal{S}_\alpha$, we increased its abundance in each subject $i$ in the treatment group by adding a random count sampled from $Binomial(N_i, Af_k)$, where $N_i$ is the sequencing depth of subject $i$, $f_k$ is the average observed proportion of taxon $k$ across all the samples in the data, $A$ controls the strength of the treatment-mediator association, and we set $A=1$. 
%The factor $A$ controls the effect size of the treatment-taxa association and we set $A=1$.  
To simulate the outcome, we used the log-contrast regression model \citep{lin2014variable} that imposes a zero-sum constraint on the association coefficients to account for the compositional nature of the covariates.
% the microbial composition covariates.
The log-contrast model was also used as the outcome model in CMM. 
In our simulation, we considered both continuous and binary outcomes.
For the continuous outcome, we generated data from the linear log-contrast regression model
\begin{equation*}
Y_i = \beta_T T_i + \sum_{k\in \mathcal{S}_\beta} {\beta_k} \log(f_{ik}) + \epsilon_i, ~~ \text{subject to}  \sum_{k\in \mathcal{S}_\beta} {\beta_k}=0,
\end{equation*}
where $f_{ik}$ is the observed proportion of taxon $k$ in subject $i$ and $\epsilon_i$ is the zero-mean normal error.  
For the binary outcome, we generated
data from the logistic log-contrast regression model
\begin{equation*}
\text{logit}\{Pr(Y_i=1)\} = \beta_T T_i + \sum_{k\in \mathcal{S}_\beta} {\beta_k} \log(f_{ik}), ~~ \text{subject to}  \sum_{k\in \mathcal{S}_\beta} {\beta_k}=0.
\end{equation*}
In these models, the coefficient $\beta_T$ was sampled from $Uniform(0, 1)$ and $\beta_k$'s were sampled from $Uniform(0, B)$, where $B$ controls the strength of the mediator-outcome association, and we set $B=1$. The values of $\beta_k$'s were centered such that the zero-sum constraint was satisfied.  
We used 2000 replicates of simulation to evaluate
empirical type I error and power of the global mediation tests
at the 0.05 significance level.

To evaluate the performance of different methods when the mediation signals are denser and the mediating taxa are randomly scattered rather than clustered on the tree, we conducted another set of simulations. Specifically, we considered $3$, $6$, and $15$ randomly scattered mediating taxa (taxa in $\mathcal{S}_\alpha$ and $\mathcal{S}_\beta$ are completely overlapped). We also lowered the mediation effect size by decreasing A to $0.001$ and $B$ to $0.5$ to better see the separation of different methods.

In addition to the global mediation test, we also evaluated the empirical FDR and power in identifying mediating nodes on the tree. All the ancestor nodes of the mediating taxa are mediating nodes. To evaluate power, we focused on the discovery rate of identifying the most recent common ancestor nodes of the mediating taxa because the signal on the upper-level ancestor nodes was diluted by the non-mediating taxa, therefore, difficult to detect.
Besides the mixture-distribution-based test implemented in PhyloMed, we performed standard Sobel's test and joint significance test to obtain the $p$-value for testing the subcomposition mediation effect in each local model. We then applied the BH procedure to select significant mediating nodes while controlling FDR at the 0.05 level.

\subsection{Simulation results}
\label{s:simResult}

For the distance-based global mediation tests MedTest and MODIMA, we present their results of Jaccard distance since it has the highest power among all distances considered. For the CMM global mediation test, we present the result from the normality-based test as it is very similar to the one from the bootstrap test. 
The current implementation of CMM cannot handle binary outcomes and the situation where the number of taxa is larger than the sample size. Therefore, the CMM result is only available for the continuous outcome and large sample size  $n=200$.  

The empirical type I error of different global mediation tests is provided in Table \ref{tab:typeI}.  
%The quantile-quantile plots of the 1000 test $p$-values of different tests under the null are shown in Figure \ref{fig:type1qq}. 
The type I error of PhyloMed is properly controlled and closer to the significance level of 0.05 than the other methods in almost all scenarios. 
The permutation version of PhyloMed (PhyloMed.P) controls type I error better than the asymptotic version (PhyloMed.A), especially when the sample size is small ($n=50$). 
MedTest and MODIMA tests are generally conservative. When $|\mathcal{S}_\alpha|=|\mathcal{S}_\beta|=0$ (i.e. all taxa are under $H_{00}$), their conservativeness is worse than other scenarios where some taxa are under $H_{10}$ or $H_{01}$. 
In contrast, the performance of the PhyloMed global test is less affected by the mixture proportions of different nulls.
%although there is a slight inflation under $H_{10}$, which is likely due to the approximation of the power function probabilities. 
The CMM test is too liberal in our simulation study, probably because its mediator model does not adequately fit the full microbial composition.  
%microbial composition model does not reflect all features of the microbiome data structure. 

\begin{table}
	\centering
	\caption{Empirical type I error of different global mediation tests.}
	\label{tab:typeI}
	\begin{tabular}{@{}lccccccc@{}}
		\toprule
		$n$ & $|\mathcal{S}_{\alpha}|$ & $|\mathcal{S}_{\beta}|$ & \textbf{PhyloMed.A} & \textbf{PhyloMed.P} & \textbf{MedTest} & \textbf{MODIMA} & \textbf{CMM} \\
		\midrule
		\multicolumn{8}{l}{\textbf{Continuous outcome}}\\
		  50 & 0 & 0 & 0.020  & 0.027 & 0.004 & 0.005 & - \\
		  50 & 3 & 0 &  0.028  & 0.034 & 0.026 &  0.027 & - \\
		  50 & 6 & 0 &  0.033 & 0.040 &  0.053 & 0.041 &  - \\
		  50 & 0 & 3 &  0.032 & 0.039 &  0.021 & 0.011 & - \\
		  50 & 0 & 6 &  0.026 & 0.033 &  0.035 & 0.019 & - \\
		    
		  200 & 0 & 0 & 0.023 & 0.027 & 0.005 & 0.004 &  0.309\\
		  200 & 3 & 0 &  0.039 & 0.042 &  0.035 & 0.040 & 0.480 \\
		  200 & 6 & 0 &  0.036 & 0.038 &  0.052 & 0.055 & 0.504 \\
		  200 & 0 & 3 &  0.039 & 0.042 &  0.035 & 0.029 &  0.149 \\
		  200 & 0 & 6 &  0.044 & 0.045 &  0.045  & 0.042 & 0.091\\		
		
		\midrule
		\multicolumn{8}{l}{\textbf{Binary outcome}} \\
		 50 & 0 & 0  & 0.022 & 0.026 & 0.003 & 0.002 & - \\
		 50 & 3 & 0  & 0.028  & 0.041 & 0.033 &  0.016 & - \\
		 50 & 6 & 0  & 0.029 & 0.035 &  0.046 & 0.033 &  - \\
		 50 & 0 & 3 & 0.025 & 0.035 &  0.013 & 0.006 & - \\
		 50 & 0 & 6  & 0.026 & 0.041 &  0.014 & 0.009 & - \\
		 
		 200 & 0 & 0  & 0.021 & 0.024 & 0.002 & 0.004 &  -\\
		 200 & 3 & 0  & 0.037 & 0.039 &  0.042 & 0.036 & - \\
		 200 & 6 & 0 & 0.037 & 0.042 &  0.049 & 0.046 & - \\
		 200 & 0 & 3 & 0.038 & 0.038 &  0.019 & 0.010 & - \\
		 200 & 0 & 6  & 0.031 & 0.032 &  0.034 &  0.020 & - \\		
		
		\bottomrule
	\end{tabular}
	\bigskip
\end{table}

\begin{figure}
	\centering
	\includegraphics[width=0.9\textwidth]{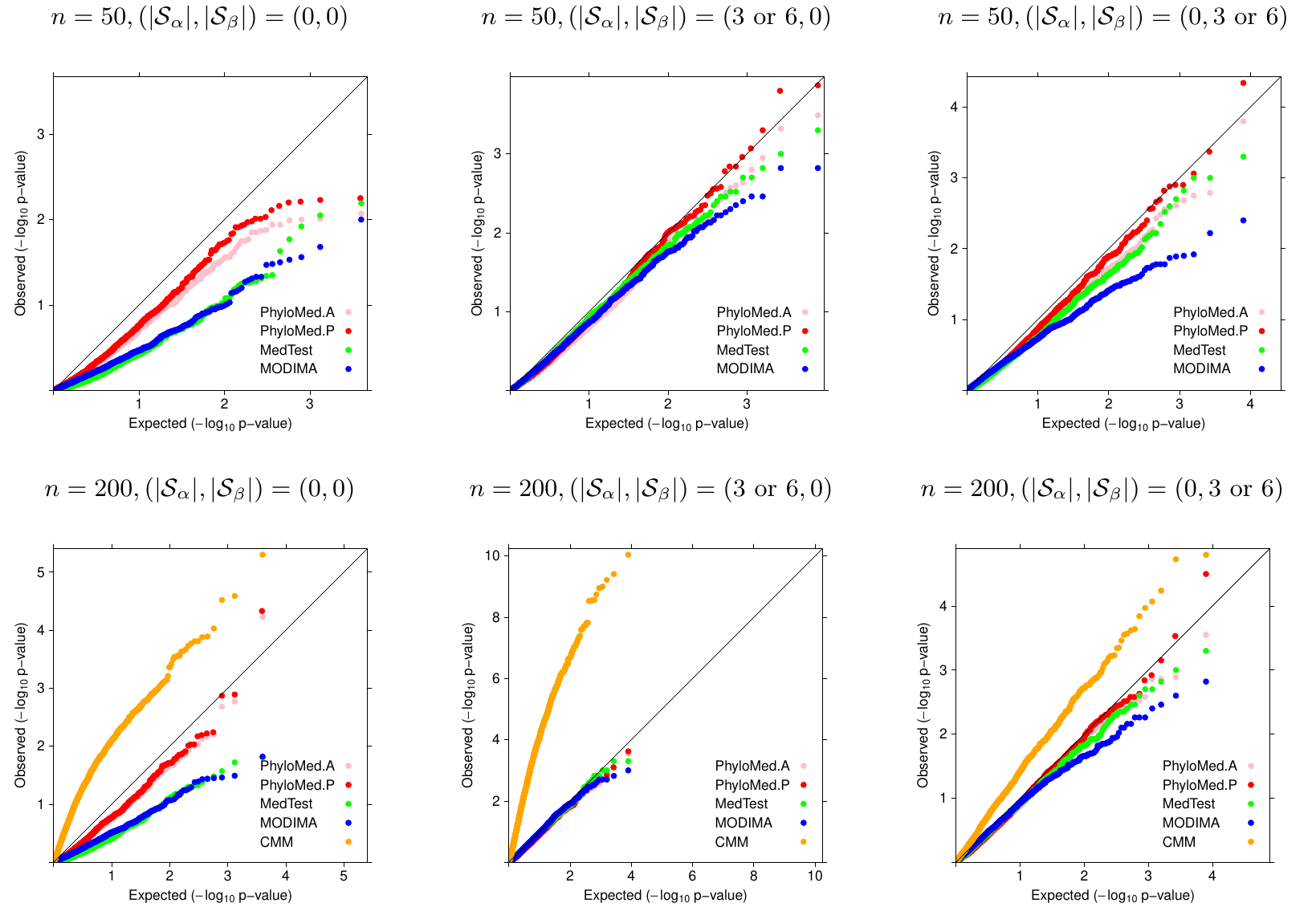}
	\caption{Quantile-quantile plots of different global mediation tests in the simulation study for the continuous outcome.}
	\label{fig:type1qq}
\end{figure}

In Figure \ref{fig:type1qq} and Web Figure S2, the control of the type I error is also
demonstrated by the quantile-quantile plots
of $p$-values from different tests under various null hypothesis settings. The PhyloMed global test yields $p$-values that are mostly aligned with
the diagonal line, whereas 
MedTest and MODIMA tests are deflated and CMM test is inflated. 

%The quantile-quantile plots of the 1000 test $p$-values of different tests under the null are shown in Figure \ref{fig:type1qq}. 

\begin{figure}
	\centering
	\includegraphics[width=0.8\textwidth]{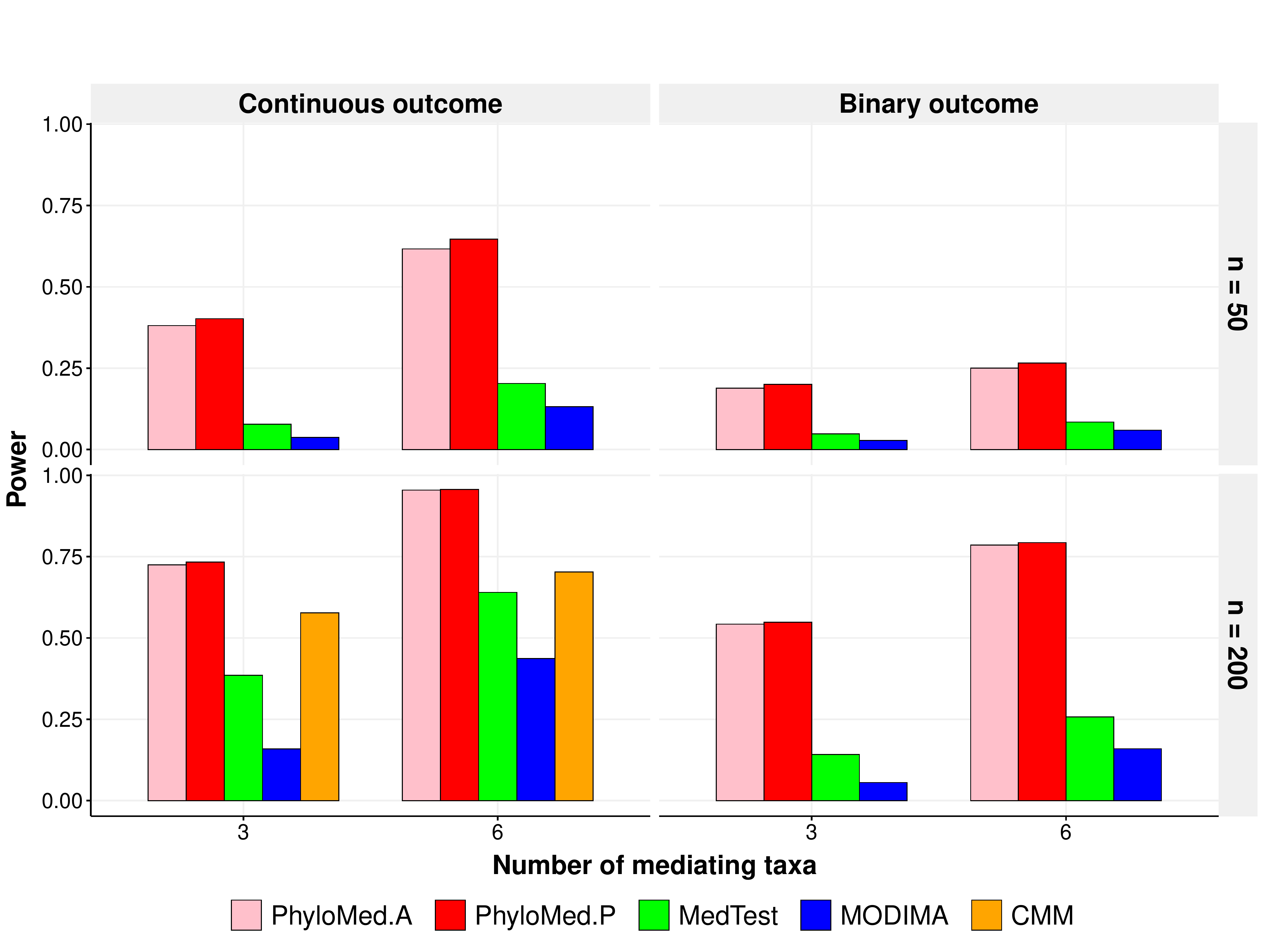}
	
	\caption{Power comparison of different global mediation tests when the treatment-associated taxa set and the outcome-associated taxa set are completely overlapped. The mediating taxa are clustered on the tree.
	}
	\label{fig:power}
\end{figure}

Figure \ref{fig:power} displays the power results for the setting where there are 3 or 6 mediating taxa clustered on the tree. 
PhyloMed is substantially more powerful than the MedTest and MODIMA in all settings. 
%The power gain can be as high as 50\%. 
PhyloMed is also more powerful than CMM even though CMM's power is overestimated due to the inflation of the type I error. 
The power gain is because PhyloMed
tests subcompositions on the ancestor nodes of mediating taxa, at which mediation signals were condensed.  
Moreover, PhyloMed employs the mixture distribution in testing the subcomposition mediation effect, which is more efficient than traditional mediation tests. The conclusion remains the same when taxa in $\mathcal{S}_\alpha$ and $\mathcal{S}_\beta$ are partially overlapped (Web Figure S3).

Web Figure S4 displays the power results when the mediating taxa are randomly scattered. When the mediation effect is large, PhyloMed is more powerful than other methods. When the effect is small, the distance-based methods can be more powerful than PhyloMed as mediation signals become denser under the small sample size (upper right panel of Web Figure S4). 
%properly accounts for the the proposed procedure accounts for
%the finite-sample mixture null distribution 
%by testing subcompositions on internal nodes
% and the power gain can be as high as 50\%. In particular, the proposed method have greatly increased power when the causal species tend to cluster on the tree.

In the procedure of detecting mediating nodes, the FDR is controlled for all mediation tests applied to local models (Web Table S1).
Web Table S2 shows the discovery rate of the most recent common ancestor nodes of the mediating taxa. PhyloMed that is based on the mixture distribution has much higher power than the standard Sobel's and joint significance tests in identifying those mediating nodes.

\section{Real Data Application}
\label{s:dataApp}

In the United States, the largest use of antibiotics is within farms, with low doses fed to farm animals to stimulate weight gain. However, there is a growing concern that antibiotic exposure may have long-term consequences.
Studies have shown that antibiotics can have a great impact on the abundance of bacteria in the gut community. 
It is interesting to investigate whether the subtherapeutic antibiotic treatment effect on body weight is mediated through the perturbation of the gut microbiome and study the underlying mechanisms. \cite{cho2012antibiotics} conducted an experiment by administrating different low-dose antibiotic treatments to young mice and evaluated changes in body fat and compositions of the microbiome in cecal and fecal samples. Specifically, mice were randomly assigned to five treatment groups (four types of antibiotics + no antibiotics), with 10 mice in each group. The body fat percentage was measured and used as the outcome in our analysis. 
The microbiome composition was established by
16S rRNA gene sequencing. The sequencing reads were preprocessed using the QIIME pipeline \citep{caporaso2010qiime} and clustered at a 97\% similarity threshold, resulting in 6547 unique microbial taxa. The phylogenetic tree for these taxa was also generated from the pipeline.   

In our analysis, the mediators include the top 100 most abundant taxa that make up 80\% and 87\% of the cecal and fecal microbial compositions, respectively. 
The sequencing read counts assigned to these taxa were
transformed into proportions after adding 0.5 to the count matrix, which is a common practice in compositional data analysis to avoid zeros in log-ratio transformation \citep{aitchison1986statistical}.
We combined samples from the four antibiotics groups and compared them with samples from the no antibiotics group (controls). After removing samples with low sequencing depth, we had 48 samples (38 in antibiotics vs 10 in controls) for cecal data analysis and 46  samples (36 in antibiotics vs 10 in controls) for fecal data analysis. 
No confounders were included in the mediation model since we assume they had been well-controlled in the randomized experiment.

The mice in the antibiotic group have a higher average body fat percentage than those in the control group (Web Figure S5). 
We used PhyloMed to investigate if the cecal and fecal microbiome mediates the effect of the antibiotics on body fat increase. We used permutation to obtain $p_{\alpha_j}$'s  and $p_{\beta_j}$'s in local models due to the small sample size of the study. 
Among all the local models in PhyloMed, the estimated proportions for different nulls $H_{00}$, $H_{10}$, and $H_{01}$ are $\widehat{\pi}_{00} = 0.65$, $\widehat{\pi}_{10}=0.30$ and $\widehat{\pi}_{01}=0.05$ for cecal data, and $\widehat{\pi}_{00} = 0.88$, $\widehat{\pi}_{10}=0.12$ and $\widehat{\pi}_{01}=0$ for fecal data. 
The PhyloMed global mediation test gives $p$-values of $0.088$ and $0.078$ for cecal and fecal data analyses. 

\begin{figure}
	\centering
	\includegraphics[width=0.7\textwidth]{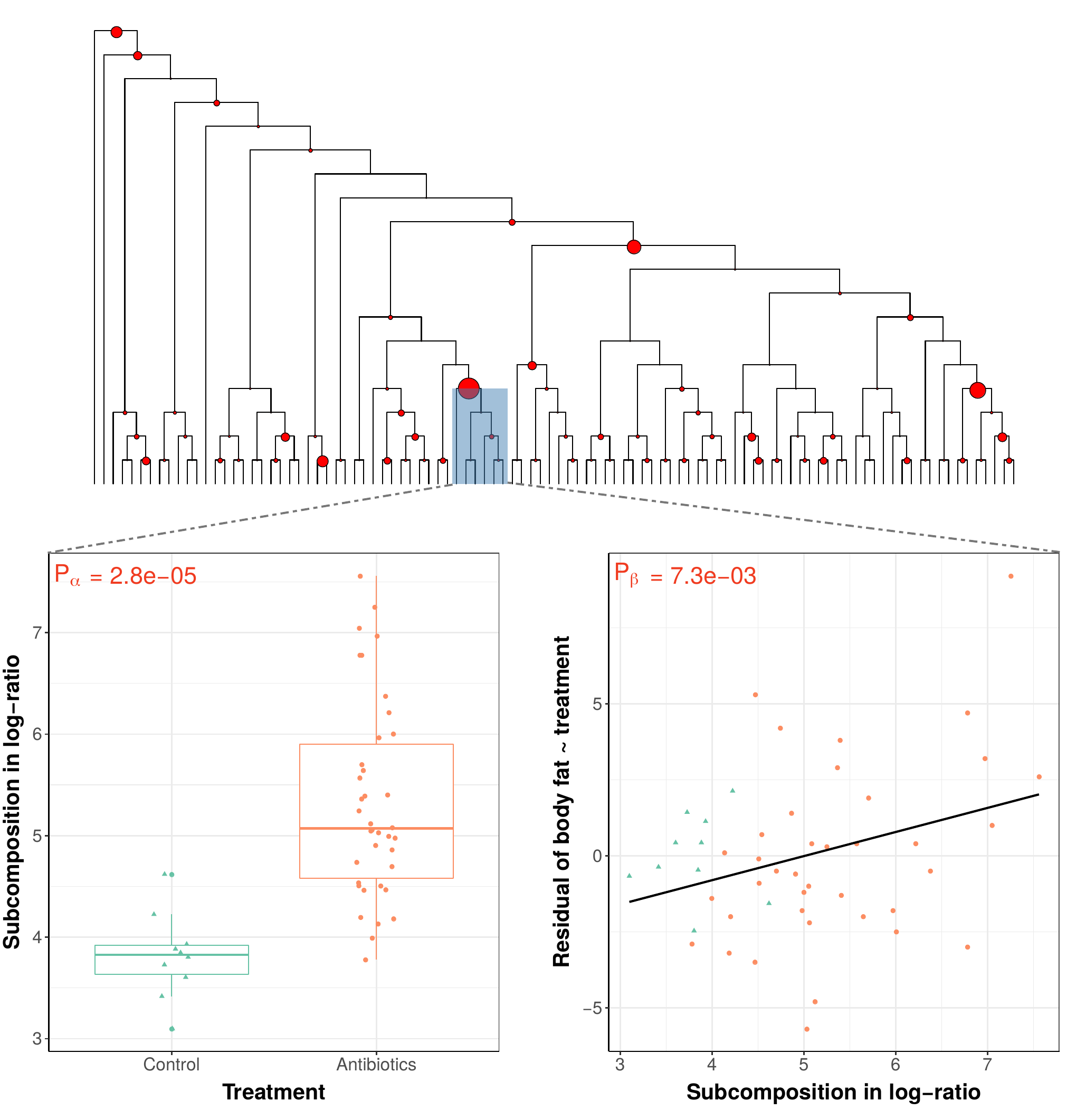}
	\caption{Identified mediating node on the phylogenetic tree and the corresponding subcomposition associations with the treatment and the body fat outcome in cecal data analysis. 
		%Cecal microbiome mediated responses to antibiotic treatment. 
		Top: phylogenetic tree, the size of the circle at each internal node is proportional to $-\log_{10}$(local subcomposition mediation test $p$-value) and the clade under the identified node is highlighted in a blue rectangle; Bottom left: association between the treatment and the log-ratio transformed subcomposition mediator; Bottom right: association between the log-ratio transformed subcomposition mediator and the body fat percentage residual after adjusting for treatment effect.
	}
	\label{fig:cecaltree}
\end{figure}

We then used PhyloMed to identify internal nodes with significant mediation effects on the phylogenetic tree while controlling FDR at 0.1. Figure \ref{fig:cecaltree} visualizes the identified mediating nodes in cecal data analysis and the relevant treatment-subcomposition and subcomposition-outcome associations. The log-ratio of subcomposition at the identified node is significantly associated with the treatment and the outcome ($p_{\alpha_j} = 2.8\times 10^{-5}$ and $p_{\beta_j} =7.3 \times 10^{-3}$). The descendants of the mediating node are a group of evolutionarily close taxa that likely share similar biological functions and jointly contribute to the mechanism of antibiotics' effect on body fat change. 
The results for the fecal data analysis were visualized similarly in Web Figure S6. 
When we employed Sobel's test and joint significance test in local models, the subcomposition mediation test $p$-values were much less significant than those in PhyloMed (Web Figure S7) and no mediating nodes were identified.
%Most subcompositions have no mediation effects and their PhyloMed $p$-values are aligned well with the diagonal line in the quantile-quantile plot (Web Figure S5). 

We applied MedTest and MODIMA to the same dataset and results are not significant at the 0.1 level for all distances (Bray-Curtis, Jaccard, weighted and unweighted UniFrac, Web Table S3).
To apply CMM, the number of taxa needs to be equal to or smaller than the sample size. We aggregated taxa along the phylogenetic tree to obtain a smaller number of taxa. None of the methods give significant $p$-values on this set of aggregated taxa (Web Table S3). 
%by merging  run CMM test,  The CMM cannot be applied here because the number of taxa is larger than the sample size. 

\section{Discussion}
\label{s:discuss}
This article has introduced a new framework PhyloMed for testing the mediation effect in high-dimensional microbial composition. The method leverages the hierarchical phylogeny relationship among different microbial taxa to decompose the complex mediation model on the full microbial composition into multiple simple independent local mediation models on subcompositions. 
This tree-guided approach effectively enriches mediation signals that tend to cluster on the phylogeny tree and boosts the power of the test for weak and sparse mediation effects among taxa. Moreover, PhyloMed properly takes into account the composite null hypothesis of the mediation test and produces well-calibrated $p$-values for testing local subcomposition mediation effects. Our simulation studies have shown that PhyloMed controls the type I error better and has substantially higher power than existing methods MedTest, MODIMA, and CMM.
Besides testing the global mediation effect of the microbial community, PhyloMed enables us to pinpoint clusters of mediating taxa at various taxonomic resolutions, represented by subcompositions with significant local mediation $p$-values. The usefulness of PhyloMed was demonstrated in the real data analysis evaluating the mediating role of the gut microbiome in the antibiotic effect on body fat.

Under the PhyloMed framework, alternative mediator models can be used and replace equation (\ref{Mmodel}). We have considered using Dirichlet-multinomial (DM) regression \citep{la2012hypothesis} and quasi-likelihood regression \citep{tang2017general} to link subcomposition count data to the treatment. The type I error results (Web Table S4) shows that the DM-based test can be too liberal and the quasi-likelihood test is more conservative than the log-ratio linear model equation (\ref{Mmodel}) adopted by PhyloMed.

The performance of PhyloMed can be affected by the accuracy of the estimates $\widehat{\pi}_{00}$, $\widehat{\pi}_{10}$, $\widehat{\pi}_{01}$. We considered an alternative approach (details in Web Table S5) to first use Storey's method \citep{storey2002direct} to obtain estimates $\widehat{\pi}_{00}$, $\widehat{\pi}_{0\bullet}$ and $\widehat{\pi}_{\bullet0}$ and then calculate $\widehat{\pi}_{01}= \widehat{\pi}_{0\bullet} - \widehat{\pi}_{00}$ and
$\widehat{\pi}_{10}= \widehat{\pi}_{\bullet 0} - \widehat{\pi}_{00}.$
This approach flexibly accommodates any distribution of (${\pi}_{00}$, ${\pi}_{01}$, ${\pi}_{10}$). 
Under our simulation setting, this approach yields similar performance as the approach described in the Method Section  (Web Table S5). 
In addition, regardless of the estimation method we used, $\widehat{\pi}_{00}$, $\widehat{\pi}_{10}$, and $\widehat{\pi}_{01}$ will generally be less accurate when the mediation signal is weak. We evaluated how this may affect the performance of PhyloMed by decreasing the mediation effect size in the simulation. 
%The estimated proportions of nulls are less accurate (Web Table S6). 
%the type I error is slightly more deflated, the and power is lower. 
%The type I error is under control, although slightly more deflated (Web Table S7). However,  the power is still significantly better than the existing methods in most of the cases (Web Figures S4 and S8).  
Although the estimated proportions of nulls are less accurate (Web Table S6) and the type I error is more deflated comparing to the large-effect-size setting (Web Table S7), the power is still significantly better than the existing methods in most scenarios (Web Figures S4 and S8).

PhyloMed is designed to detect the sparse mediation signal among taxa. 
%by decomposing full composition into subcompositions and assuming the mediation effect is driven by a small number of subcompositions. 
%the proposed global test of mediation is more powerful than PERMANOVA
%type of tests when the overall composition difference is due to a few subcompositions since our test considers each  subcomposition separately and then combines the
%$p$-values. 
Since 
all $p$-values across local mediation models are independent under the mediation null hypothesis, many standard $p$-value combination methods and FDR-controlling procedures can be employed.  
In this article, we focus on Simes's $p$-value combination that is sensitive to detect the sparse signal. If the mediation signal is dense (i.e. many taxa in the microbial community have mediation effects), alternative methods such as Fisher's and harmonic mean $p$-value combination methods \citep{wilson2019harmonic} may yield better power.    
In addition, there are FDR-controlling procedures that leverage the hierarchical tree structures \citep{benjamini2003hierarchical, li2019multiple, lei2021general} and potentially have superior performance when mediation hypotheses defined on the nearby or nested clades are more likely to be jointly true or false.  
Evaluating these different methods is beyond the scope of this paper.

In this article, we focused on using a known binary phylogenetic tree with each subcomposition on the internal node consists of two components. The method can be extended to non-binary trees (e.g. taxonomic trees) that have more than two children at internal nodes. In this case, the local mediation model will have more than one log-ratio transformed mediator. We may develop a variance-component mediation test similar to the one proposed in \cite{huang2019variance} to test multivariate mediators. Furthermore, for the microbiome data derived from shotgun metagenomics sequencing, multiple phylogenetic trees can be constructed, each of which is based on sequencing data of a set of marker genes. The structures of these trees may or may not agree with each other and may be subject to various degrees of error. Evaluating the robustness of the method against tree misspecification is a future effort. 
%In the real data example, the tree is learned from 16S rRNA sequences using a fast and approximate maxi

%  This section is optional.  Here is where you will want to cite
%  grants, people who helped with the paper, etc.  But keep it short!

\section*{Acknowledgments}

This work
was supported by the NIH grant R01GM140464, NSF grant DMS-2054346, and Data Science Initiative
Award provided by the University of Wisconsin-Madison Office of the Chancellor and
the Vice Chancellor for Research and Graduate Education with funding from the Wisconsin
Alumni Research Foundation. The work was
also supported in part by the NIH/National Institute on Aging through the University of Wisconsin
Madison Center for Demography of Health and Aging Pilot Grant (P30AG017266).

\section*{Supporting Information}
A software implementation of our method as well as all code to reproduce the results in this paper is available online with the posting of this article, or on Github at 
\url{https://github.com/KiRinHong/miMediation}.

\bibliographystyle{unsrtnat}
\bibliography{references}  %%% Uncomment this line and comment out the ``thebibliography'' section below to use the external .bib file (using bibtex) .





% \end{thebibliography}

\end{document}